\documentclass[preprint,12pt]{elsarticle}
\usepackage{graphics}
\usepackage{graphicx}
\usepackage{amssymb}
\usepackage{amsthm}
\usepackage{lineno}

\journal{Nuclear Instruments and Methods in Physics Research Section A }

\begin{document}

\begin{frontmatter}
\title{First results with a microcavity plasma panel detector}
\author[label1]{R.~Ball}
\author[label3]{M.~Ben-Moshe}
\author[label3]{Y.~Benhammou}
\author[label3]{R.~Bensimon} 
\author[label1]{J.~W.~Chapman} 
\author[label3]{M.~Davies} 
\author[label3]{E.~Etzion} 
\author[label1]{C.~Ferretti} 
\author[label4]{P.~S.~Friedman}
\author[label1]{D.~S.~Levin} 
\author[label3]{{Y.~Silver}} 
\author[label2]{R.~L.~Varner}
\author[label1]{C.~Weaverdyck}
\author[label1]{B.~Zhou}
\address[label1]{ University of Michigan, Department of Physics, 
                  Ann Arbor, Michigan, 48109, USA.}
\address[label2]{ Oak Ridge National Laboratory, Physics Division, 
                  Oak Ridge, Tennessee, 737831, USA.}
\address[label3]{ Tel Aviv University,
                  School of Physics and Astronomy, Tel Aviv, Israel.}
\address[label4]{ Integrated Sensors, LLC, Ottawa Hills, Ohio, 43606, USA.}

\begin{abstract}
A new type of gaseous micropattern particle detector based on a closed-cell 
microcavity plasma panel sensor is reported. 
The first device was fabricated with $\rm 1 \times 1 \times 2 \; mm$ cells. 
It has shown very clean signals of 0.6 to 2.5 volt amplitude, fast rise 
time of approximately 2~ns and FWHM of about 2~ns with very uniform signal 
shapes across all pixels. 
From initial measurements with beta particles from a radioactive source,
a maximum pixel efficiency of greater than 95\% is calculated, for operation
of the detector over a 100V~wide span of high voltages (HV). 
Over this same HV range, the background rate per pixel was measured to be 
3 to 4 orders of magnitude lower than the rate with the cell illuminated 
by the beta source. 
Pixel-to-pixel count rate uniformity is within 3\% and stable 
within 3\% for many days. 
The time resolution is 2.4~ns, and a very low cell-to-cell crosstalk
has been measured between cells separated by 2~mm.
\end{abstract}
\begin{keyword}
\hspace{-10pt}micropattern gas detector; particle detector; plasma panel sensor.
\end{keyword}
\end{frontmatter}

\section{Introduction}\label{Introduction}

Plasma panel sensors (PPS) with {\it open}-cell structures as particle 
radiation detectors have been investigated by our group for several 
years~\cite{PDP}. 
The first results for a new radiation detector based on a {\it closed}-cell 
microcavity-PPS structure  are here described.
This research aims at developing scalable, inexpensive, low mass, 
long life and hermetically sealed gaseous detectors for both scientific
and commercial applications. 
\par
The panel acts as an array of independent closed gas pixels/cells
biased to discharge when free-electrons or ions are generated in the 
gas by ionizing radiation.
The electron avalanche is self-contained by the walls that define the 
cell itself and suppressed using Penning mixtures~\cite{Penning} with 
quenching gases and a localized resistance at each pixel.
The HV applied to each pixel is chosen such that the mode of operation 
is in the Geiger region~\cite{KNOLL}, and rendering this device 
intrinsically digital.
The cell capacitance, which was measured to be $ 0.3 \pm 0.1 $~pF, 
stores the total charge available for a pulse.

\begin{figure}[!ht]\centering
  \includegraphics[width=0.48\textwidth]{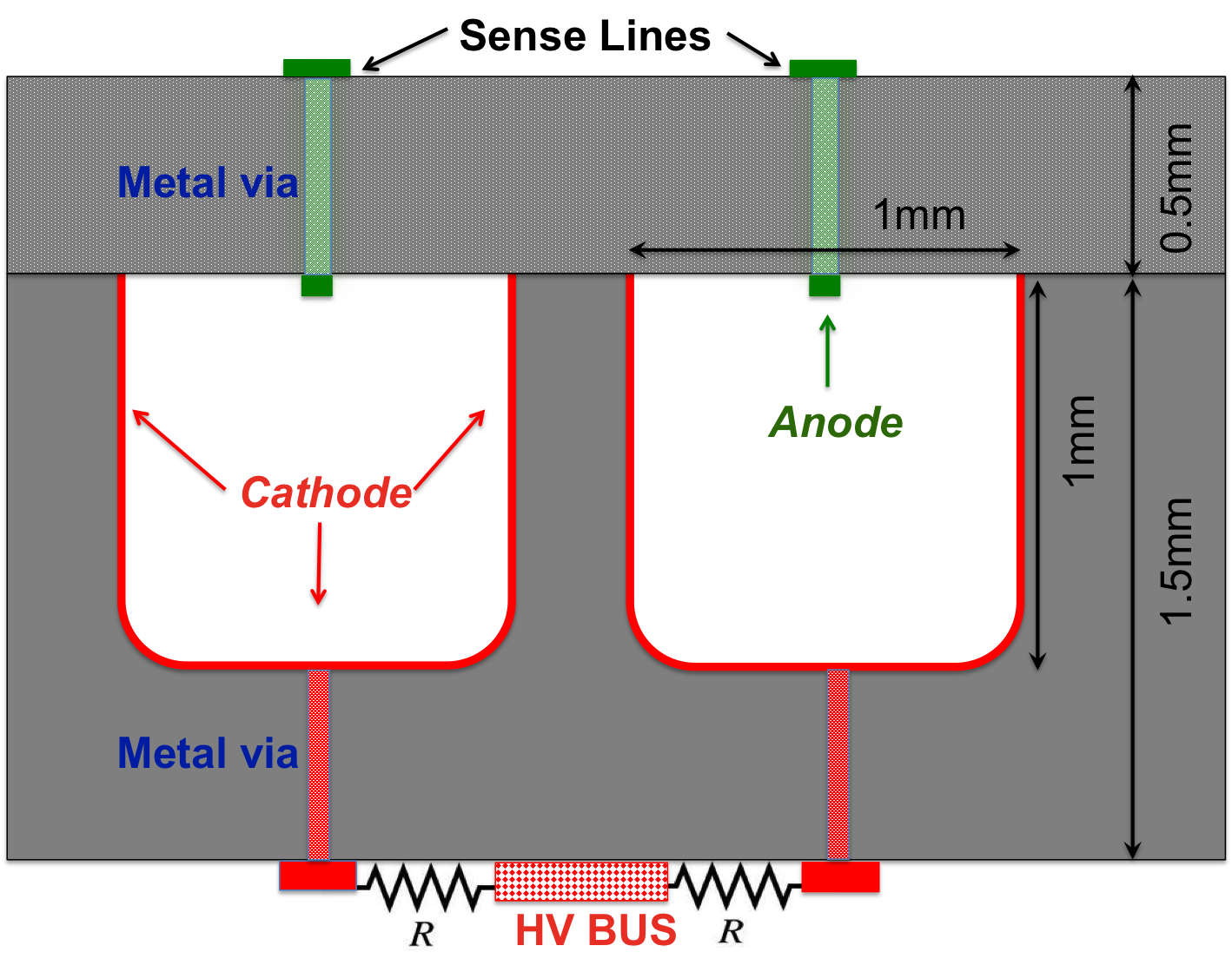}
  \vspace{-4mm}
  \caption{Concept side view of a microcavity-PPS panel.}
  \label{fig:Concept}
\end{figure}

\begin{figure}[!htb]\centering
  \includegraphics[width=0.48\textwidth]{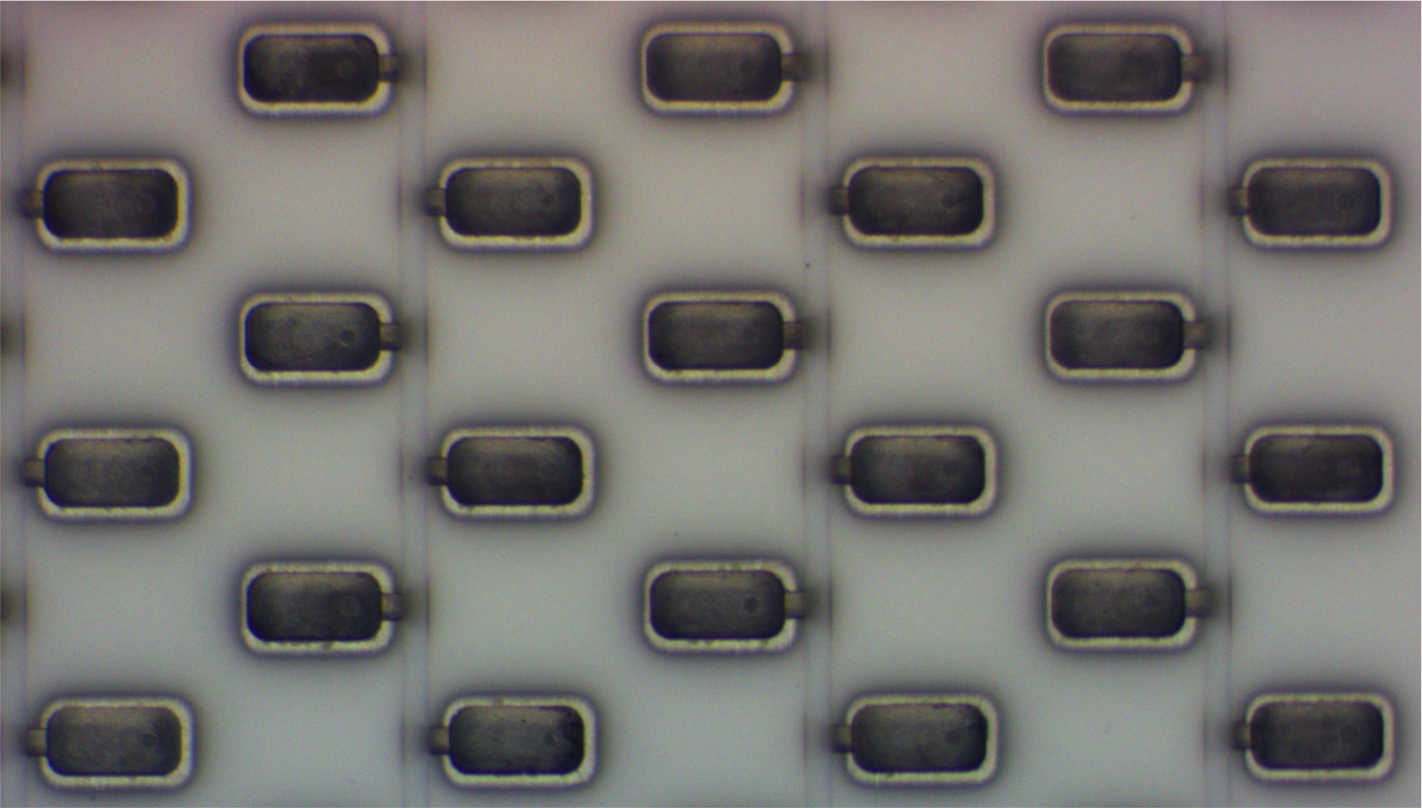}
  \vspace{-4mm}
  \caption{Photograph of part of a microcavity-PPS bottom substrate with 
           $\rm 1 \times 1 \times 2 \; mm $ metalized microcavity array. 
           Three thin gas lines (vertical) are visible, each servicing 
           two columns of cavity pixels.}
  \label{fig:Array}
\end{figure}

The detector consists of two substrates (see Fig.~\ref{fig:Concept}) sealed 
together.
The top supports the anode in each cell, which is connected by a metal via to 
a readout (RO) or sense line on the outside.
The bottom hosts the metalized cavities, each one with a metal via connecting 
the cathode to the HV distribution by means of an external quench resistor.
In this first prototype, cells of $\rm 1 \times 1 \times 2 \; mm $  
were arranged with a low packing fraction of 18\% as shown in 
Fig.~\ref{fig:Array}.
Smaller cells with much higher packing fractions are planned for the next 
generation of detectors.

\section{Experimental Results}

\subsection{Setup and DAQ}

The gas used for the experiments described here was 1 atm of Argon-Neon
Penning mixture with CF$_{4}$ added to improve the response with and without
a radioactive source.
Each pixel was instrumented with a 1G$\Omega$ quench resistor.
The response was investigated with a $^{106}$Ru $\beta$-emitter 
source.
\par
In most of the tests the readout signals were first discriminated
before being sent to a Wiener NIMbox, configured as a 20 channels scaler 
using customized firmware, readout by locally developed LabVIEW code.
\par
For efficiency and time resolution  measurements, the coincidence of an 
external pair of scintillator detectors was used to generate a trigger 
for the readout, using 20 ns wide, NIM logic signals.
The coincidence window of the panel and trigger was 2~$\mu$s. 
\par
In order to acquire the signal time spectrum, a portable version 
of the LHC ATLAS precision muon chambers readout system called 
MiniDAQ~\cite{MDTelx} is used.
This system is capable of recording integrated charge and 
times relative to the trigger with 0.78 ns precision.

\subsection{Signal response}

\begin{figure}[!ht]\centering
  \includegraphics[width=0.47\textwidth]{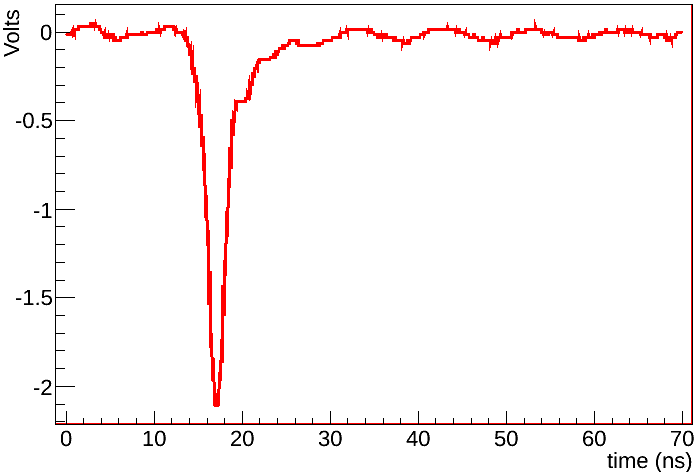}
  \vspace{-4mm}
  \caption{Example of microcavity-PPS pulse operating at 1150V.}
  \label{fig:SinglePulse}
\end{figure}

The signal produced by a discharge 
is very clean $( S/N > 20 )$, has a range of amplitudes from 600~mV 
(at HV=800V) to 2.5~V (at HV=1200V) requiring no amplification electronics, 
and has rise times of about 2~ns and FWHM of between 2 to 3~ns,
as shown in Fig.~\ref{fig:SinglePulse}. 

\begin{figure}[!ht]\centering
  \includegraphics[width=0.47\textwidth]{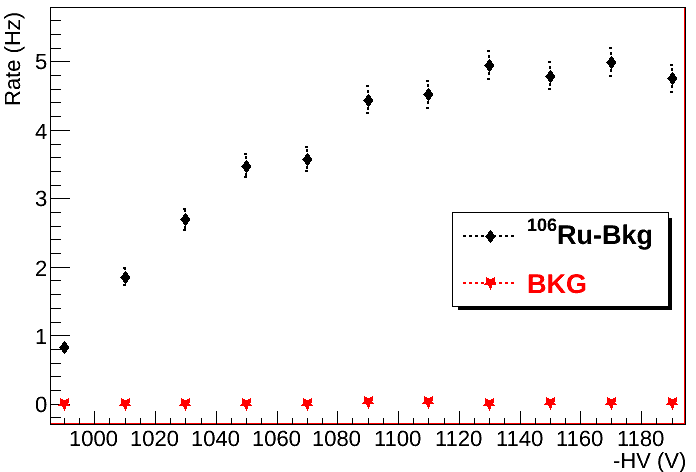}
  \vspace{-4mm}
  \caption{Single microcavity-PPS pixel rate response to $^{106}$Ru at 
           different HV.}
  \label{fig:VoltageScan}
\end{figure}

Fig.~\ref{fig:VoltageScan} shows the hit rate of a single pixel, exposed 
to the collimated source, as a function of the applied high voltage. 
A plateau between 1090V and 1190V is clearly visible.
Over the whole range of HV tested, less than 0.01 Hz of background per 
pixel were measured whereas with the source, a count rate of 2 to 3 orders 
of magnitude higher was observed. 
Data taken with an uncollimated source shows that a single pixel can count
at rates up to 400 Hz or more (at higher voltages) maintaining a very low 
background with a 3 to 4 orders of magnitude source to background ratio.

\subsection{Pixel uniformity}

\begin{figure}[!ht]\centering
  \includegraphics[width=0.47\textwidth]{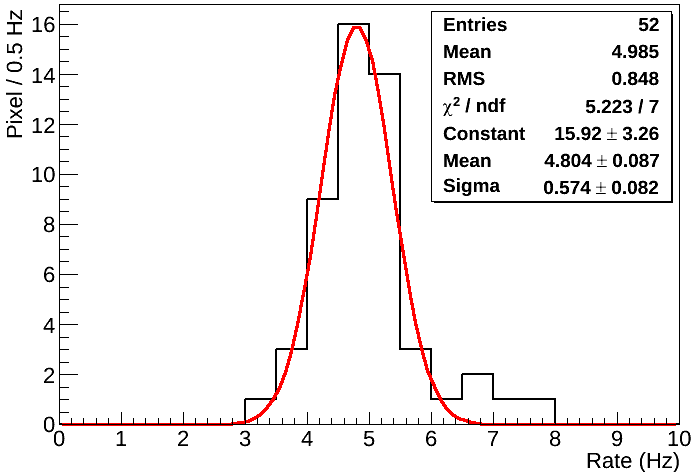}
  \vspace{-4mm}
  \caption{Histogram of microcavity-PPS single cell response rates to a 
           collimated $^{106}$Ru source at 1050V for 52 pixels.
           The distribution is fitted with a Gaussian function.}
  \label{fig:RateUniformity}
\end{figure}

The response rates of different pixels was measured to determine
the quality of the parts and their assembly.
Fig.~\ref{fig:RateUniformity} shows a histogram of the rate at fixed voltage 
of pixels illuminated by the collimated $^{106}$Ru source.
The distribution has a variance of about 12\% over 52 pixels 
tested (see Fig.~\ref{fig:RateUniformity}).

\begin{figure}[!ht]\centering
  \includegraphics[width=0.47\textwidth]{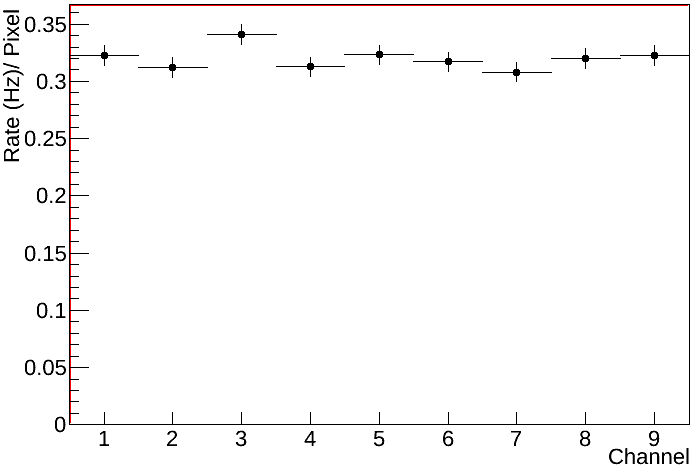}
  \vspace{-4mm}
  \caption{Microcavity-PPS response rate exposed to an uncollimated $^{106}$Ru 
           source 13 cm above the panel. The RMS on the Y-axis is 3\%.}
  \label{fig:RateUniformity2}
\end{figure}

A more precise measurement of the pixel response uniformity was obtained by
exposing the detector to a source placed at a distance far away enough to 
produce an approximate uniform flux of radiation over the whole panel.
The rate measured on each readout line was 
proportional to the 
number of pixels instrumented on that line (Fig.~\ref{fig:RateUniformity2}).
The RMS variation was about 3\%, smaller than the variance determined from 
Fig.~\ref{fig:RateUniformity}. 
The difference is largely due to alignment errors in placing the collimator 
on each pixel.

\subsection{Pixel isolation}

\begin{figure}[!ht]\centering
  \includegraphics[width=0.47\textwidth]{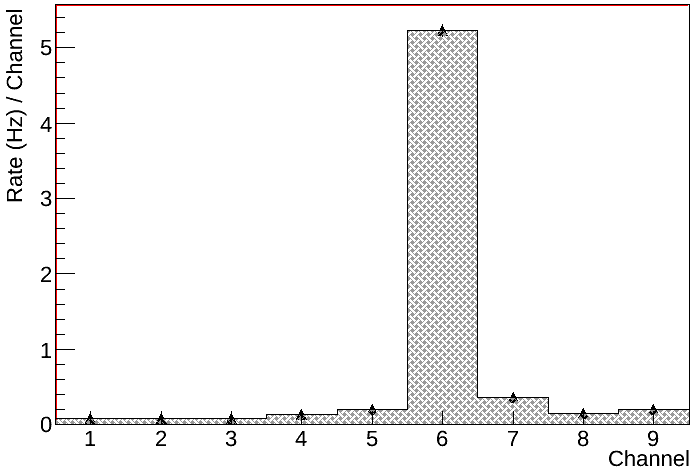}
  \vspace{-4mm}
  \caption{Hit map of the microcavity-PPS with 22 active pixels 
           (not all the lines have the same number of pixels) and where a 
           $^{106}$Ru source was collimated over a pixel on RO \#6.}
  \label{fig:Isolation}
\end{figure}

The pixel isolation was directly confirmed by the result shown in 
Fig.~\ref{fig:Isolation}. 
This plot displays the hit map associated with the collimated  $^{106}$Ru 
source placed over a single pixel in readout (RO) line \#6, in a configuration 
that had 21 other active cells nearby.

The measured rate on the illuminated line was significantly larger than the 
one on the other lines which remained very close to the background level, 
confirming that a pixel discharge on one line does not affect the others.
The slightly higher observed rate on the two lines adjacent to the 
illuminated is due to the collimator diameter being slightly larger 
than the cell, which leads to a small collateral leakage contribution 
to the neighbors.

\subsection{Time stability}

\begin{figure}[!ht]\centering
  \includegraphics[width=0.47\textwidth]{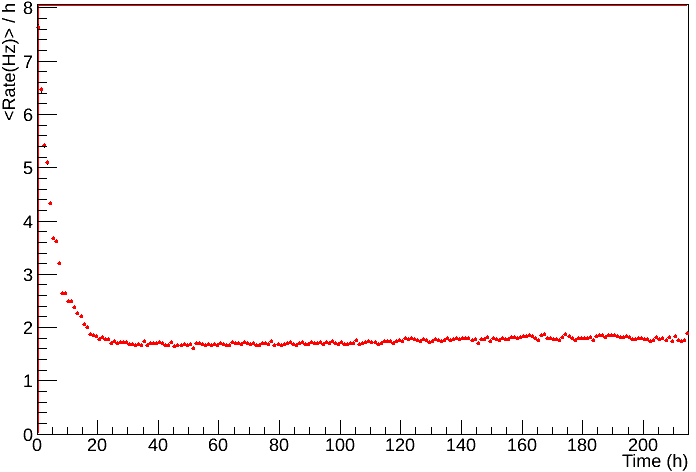}
  \vspace{-4mm}
  \caption{Microcavity-PPS single pixel response 
           as a function of time at 1000V.}
  \label{fig:Stability}
\end{figure}

Fig.~\ref{fig:Stability} exhibits the stability of the pixel response rate 
in time: the plot is the average hourly count rate under the uncollimated 
$^{106}$Ru source placed 15 cm above the panel. 
The first 18-20 hours show a fast change in response rate due to the 
individual gas components still mixing (i.e., the DAQ was 
started immediately after filling the panel one gas component at the time).
After this period, the RMS variation was $\sim $3\% over the next 9 days.

\subsection{Efficiency}

The pixel efficiency is defined as the ratio of the 
number of events with a coincidence between a pulse in the microcavity 
pixel and a trigger, divided by the total number of triggers.
The calculation of the efficiency was corrected using GEANT4~\cite{GEANT}
mainly because of two effects that increased the number of triggers:
1) the collimator diameter was larger than 1 mm, causing a fraction
   of the source $\beta$'s to trigger an event even when not passing 
   through the pixel.
2) triggers due to secondary electrons from X-rays produced by the 
   $^{106}$Ru $\beta$'s interacting either in the panel substrates 
   or in the scintillators.
The trigger rate was also measured with a plate identical to the collimator 
but without the hole: roughly one third of the triggers are recorded in 
this configuration.
This data driven method of subtracting the fraction of triggers coming 
from the source and not going through the pixel active area reduced the 
dependence of the result on simulation.

\begin{figure}[!ht]\centering
  \includegraphics[width=0.47\textwidth]{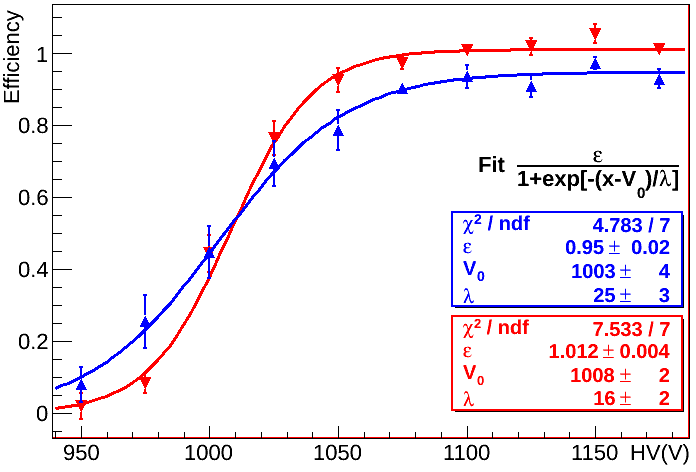}
  \vspace{-4mm}
  \caption{Microcavity-PPS pixel efficiency as a function HV
           measured for two pixels.}
  \label{fig:Efficiency}
\end{figure}

Fig.~\ref{fig:Efficiency} shows the pixel efficiency measured as a function 
of the applied HV for two pixels. 
The fit lines, based on a Fermi-Dirac function, each show a plateau, one 
at $ \varepsilon \sim 95\%$ and another at $ \varepsilon \sim 100\%$.
The total systematic error was estimated to be about 10\%, and was due to a
misalignment between the collimated source and the pixel, and the pixel 
relative to the trigger.

\subsection{Time resolution}

\begin{figure}[!ht]\centering
  \includegraphics[width=0.47\textwidth]{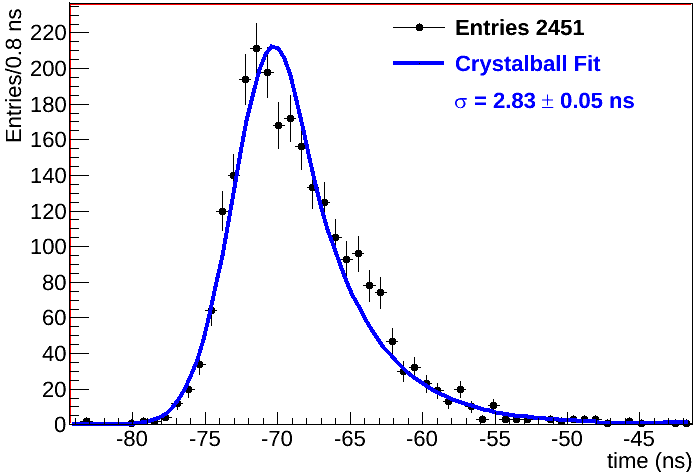}
  \vspace{-4mm}
  \caption{Distribution of the microcavity-PPS single cell arrival time 
           at 1530V.}
  \label{fig:TimeResponse}
\end{figure}

The response time of single pixel hits from the collimated $^{106}$Ru 
source is shown in Fig.~\ref{fig:TimeResponse}, where the number of
hits is plotted as a function of the pulse arrival time after 
subtraction (event by event) of the trigger time.

The experimental points were fitted with a Crystal Ball function, a 
convolution of a Gaussian with a power law accounting for an asymmetric 
tail~\cite{CBfunction}.
The Gaussian part of the distribution reflects the stochastic nature in
which the avalanche leads to the discharge,
while the tail represents the longer drift times of electrons arriving 
from regions with lower field and farther away from the anode.
The time resolution, defined as the dispersion in the time difference between
the pixel signal threshold crossing and the trigger time, shown by the Gaussian 
sigma of the fit is $\sim 2.8$~ns.
This width includes the trigger jitter, measured at 1.5 ns. 
The approximate procedure of subtraction in quadruture of the trigger 
component yields a Gaussian time resolution of 2.4 ns.
The distribution's negative mean was due to a lack of correction for the 
trigger delay with respect to the pulse (e.g. extra cables, electronics, 
etc.).

\section{Summary}

The microcavity-PPS detector prototype showed very promising results
in terms of pixel-to-pixel uniformity and time-stability of both signal 
shape and rates. The prototype has also demonstrated very low background 
over a wide range of applied high voltages, excellent pixel response 
isolation, time resolutions of a few nanoseconds, and efficiencies above 
95\% over a 100~volt range for beta particles emitted by a $^{106}$Ru 
radioactive source.
\par
The geometrical parameters of the panel microcavity cell design can be 
tuned for specific applications including smaller cells and much higher
packing fractions.
These changes, together with the optimization of the gas mixture and 
reduced quench resistances, could increase the maximum response rate,
improve both the spatial and time resolution, and increase the overall 
efficiency.
Based on these initial results, the development of this technology should 
provide a promising alternative for particle detection.

\section*{Acknowledgements}

Development of the PPS project was funded by the U.S. Department of Energy 
(DOE) - Office of Nuclear Physics Small Business Innovation Research grant
award numbers DE-SC0006204 and DE-FG02-07ER84749 to Integrated Sensors, LLC;
U.S. DOE, Office of Nuclear Physics, Applications of Nuclear Science and 
Technology grant to Oak Ridge National Laboratory, operated by UT-Battelle,
LLC for the U.S. DOE;
and DOE - Office of High Energy Physics grant number DE-FG02-12ER41788 to 
the University of Michigan.  
The research at Tel Aviv University was supported by the I-CORE Program of
the Planning and Budgeting Committee and the Israel Science Foundation 
(grant number 1937\textbackslash 12). 
Funding for scientific exchange and collaboration between Tel Aviv University 
and the University of Michigan was provided by the Israel-American Binational
Science Foundation, grant number 1008123.
\par
Disclaimer:  This report was prepared as an account of work sponsored by 
an agency of the United States Government.  
Neither the United States Government nor any agency thereof, nor any of 
their employees, makes any warranty, express or implied, or assumes any 
legal liability or responsibility for the accuracy, completeness, or 
usefulness of any information, apparatus, product, or process disclosed, 
or represents that its use would not infringe privately owned rights.
Reference herein to any specific commercial product, process, or service 
by trade name, trademark, manufacturer, or otherwise does not necessarily 
constitute or imply its endorsement, recommendation, or favoring by the 
United States Government or any agency thereof.
The views and opinions of authors expressed herein do not necessarily state 
or reflect those of the United States Government or any agency thereof.

\end{document}